\documentclass[fleqn,usenatbib]{mnras}

\usepackage[T1]{fontenc}

\DeclareRobustCommand{\VAN}[3]{#2}
\let\VANthebibliography\thebibliography
\def\thebibliography{\DeclareRobustCommand{\VAN}[3]{##3}\VANthebibliography}

\usepackage{graphicx} 
\usepackage{amsmath}  
\usepackage{amssymb}  
\usepackage{algorithmic}
\usepackage{pdflscape}
\usepackage{caption}
\usepackage{subcaption}
\usepackage{color,soul}
\usepackage{afterpage}

\title[Constraining Tidal Dissipation from Binary Star Spin]{Constraining Tidal Quality Factor using Spin Period in Eclipsing Binaries}

\author[Ruskin Patel and Kaloyan Penev]{Ruskin Patel$^{1}$ and Kaloyan Penev$^{1}$
\thanks{e-mail: rxp163130@utdallas.edu}
\\
University of Texas at Dallas
}

\date{Accepted XXX. Received YYY; in original form ZZZ}

\pubyear{2021}

\begin{document}
\label{firstpage}
\pagerange{\pageref{firstpage}--\pageref{lastpage}}
\maketitle

\begin{abstract}
Evolution of binary objects under the influence of tides drastically affects the expected observational properties of the system. With the discovery of a large number of close-in hot Jupiter systems and eclipsing binaries from missions such as \textit{Kepler} and \textit{TESS}, it has become imperative to understand the extent of tidal influence on their formation and observed properties.  In the case of binary systems, an efficient tidal dissipation can lead to either spin up or spin down of the stars and/or spin-orbit synchronization, depending upon the exchange of angular momentum between the star and the orbit. We combine the eclipsing binary systems from the \textit{Kepler} mission with stellar and orbital parameters available in the literature to create a catalog of 41 eclipsing binaries suitable for analysis of tidal dissipation. Empirically, the efficiency of tidal dissipation is parameterized using a modified Tidal Quality Factor($Q_{\star}^{'}$). We find constraints on $Q_{\star}^{'}$ using the observed rotation period of the primary star in the eclipsing binary systems. We calculate detailed evolutions of binary systems under the combined influence of tides, stellar evolution, and loss of stellar angular momentum to magnetic winds, and perform Markov Chain Monte Carlo simulations to account for the uncertainties in the observed data. Our analysis shows that $\log_{10}{Q^{'}_{\star}}=7.818\pm0.035$ can reproduce the observed primary star spin in almost all systems in our sample.
\end{abstract}




\section{Introduction}\label{sec:1}
Tidal interactions in binary systems play a pivotal role in determining the observed orbital and stellar properties. The discoveries of hot-Jupiters with tight orbits have rekindled the interest in understanding the dissipation of energy in stars through tidal interactions (\cite{Dawson_2018}). 
While some studies suggest that the physical mechanism responsible for the tidal dissipation is the turbulent cascade of energy from large to small scales (\cite{Zahn_89,Zahn_Bouchet_89,Goldreich_Keeley_77,Penev_et_al_09}), other suggest the damping of internal waves inside the star(\cite{Goodman_Dickson_98,Essick_Weinberg_16,Ogilvie_13,Rieutord_Valdettaro_10}). With several flavors of each type of mechanism suggested, the exact dissipation mechanisms are still under debate, leading to inconsistencies among various models (\cite{Ogilvie_2012,Vidal_2020,Preece_2019}). 
The efficiency of the tidal dissipation has also been studied empirically, often parameterized as a dimensionless parameter: tidal quality factor $Q^{'}_{\star} = Q_{\star}/k_2$ (\cite{Goldreich_Soter_66}), where $1/Q_{\star}$ is the energy lost over one tidal period normalized by the energy stored in the tidal deformation; $k_2$ is the ratio between the linear perturbation of the self-gravitation potential induced by the presence of the companion and the perturbing tidal potential evaluated at the stellar surface. Since many of the binary parameters (orbital period, eccentricity, obliquity, stellar/planetary rotation period) depend upon the efficiency of tidal dissipation, the parameters of the $Q^{'}_{\star} $ model can be inferred using observations alone.

Observational studies using different methods for constraining $Q_{\star}$ have shown significant inconsistencies among themselves. \citet{Meibom_2006} used the tidal circularization method on solar-type binaries in open clusters to find the tidal circularization period. Their work gives an approximate constraint of  $Q_{\star}^{'} \sim 10^{6.0}$. \citet{Jackson_2007}  assumed that close-in exoplanets started with similar eccentricity distribution as longer period ones. The authors integrated the tidal evolution equations backward in time to obtain initial eccentricity and orbital period. The best fit for their model gives a $Q_{\star}^{'} \sim 10^{5.5}$. On the other hand, \citet{Hansen_2010} adopted a simplified tidal evolution model and an alternative dissipation parameterization to obtain constraints between $10^{7.5}-10^{8.5}$. \citet{Bonomo_2017} analyzed 231 transiting giant planets and used a model similar to the one described in \citet{Goldreich_Soter_66} to obtain lower limits on $Q_{\star}$ between $10^4 - 10^8$. \citet{Penev_2018} used rotation periods of stars hosting hot Jupiters to obtain $Q_{\star}$ changing gradually with tidal period in the range $10^5-10^7$.  Although most of these studies have been conducted mainly on star-planet systems, robust constraints on $Q_{\star}^{'}$ can also be obtained by studying the tidal effects in star-star systems.

There are  multiple physical mechanisms proposed for tidal dissipation throughout the literature, each predicting different $Q_{\star}$. \cite{osti_4471673,Zahn_89,Goldreich_Keeley_77} first gave the theory of equilibrium tide under which the gradient of gravitational potential over the stars in binary system gives rise to physical deformation, perturbing the hydrostatic equilibrium in the star. This creates a large-scale fluid displacement and the dissipation is through the friction applied by turbulence in the star \citep{Zahn_89,Zahn_2008,duguid2021}. On the other hand, under the dynamical tidal theory,  the tidal forces can also give rise to inertial and gravity waves with Coriolis force and buoyancy as the restoring forces inside the stars. For low mass stars with convective envelope and radiative core, the inertial waves can reflect by the radiative core and stellar surface resulting in enhanced dissipation \citep{Ogilvie_13}.

The internal gravity waves generated in the radiative zone of the stars contribute to the tidal dissipation depending upon the internal structure of the star. In low mass stars with radiative core and convective envelopes, the internal gravity waves launched  at the boundary of radiative and convective zones, propagate inwards through the radiative core. The resonance between the stellar oscillation modes and the tidal frequencies can enhance the tidal dissipation. \cite{Fuller_2017} explores the theory of resonance locking, based on  linear analysis of the dynamical tides. Resonance locking predicts $Q^{'}_{\star} \propto P_{tide}^{-13/3}$ \citep{Ma2021OrbitalDO}. On the other hand, the damping of non-linear g mode waves can also lead to the dissipation in the radiative zone of the star. Waves get excited at the boundary of convective and radiative zones and propagate inwards towards the center, where their amplitude is amplified. If the waves are highly non-linear, instead of resonance locking, wave breaking occurs near the center of the star. This results in formation of travelling g mode waves, instead of standing g modes and resonance locking cannot occur \citep{barker_ogilvie_2011} and the tidal quality factor has a dependence $Q^{'}_{\star} \propto P_{tide}^{2.6}$. \cite{Essick_Weinberg_16} has also examined  non-linear damping of g modes excited by hot Jupiters. According to the authors, in case of weak non-linearity, these g waves would not break but further lead to excitation of daughter and grand-daughter modes that dissipate energy. The authors provide best fit to their model with $Q \propto P_{tide}^{2.4}$. For current analysis,  we assumed a constant value for the $Q^{'}_{\star}$, we will explore the dependencies with detailed frequency dependent formulation of $Q^{'}_{\star}$ in future.

This paper derives tight constraints on $Q_{\star}^{'}$ using the observed rotation periods of the primary stars in eclipsing binaries. A simplified explanation of the method is as follows. Under the equilibrium tide theory, the gradient of gravitational potential over the extended objects in a binary system gives rise to deformation in these objects. The deformed body either lags behind or ahead of the gravitational potential of the secondary due to tidal dissipation, giving rise to angular momentum exchange between the stars and the orbit. Depending upon the exchange of angular momentum, the stars can either spin up or spin down. For instance, if the primary star (more massive star in the binary) is rotating slower than the orbit, the tides spin up the star while the orbital period goes down and vice-versa. This continues until a state of synchronization is achieved where the rotation period of the star is equal to the orbital period. Further evolution is very slow and only happens because of the loss of angular momentum due to magnetic winds, assuming the binary system is isolated. Hence, we can say that the observed rotation periods result from a long history of tidal interaction in binary systems and can be used as a tool to derive constraints on tidal dissipation or $Q^{'}_{\star}$.

In order to simulate the combined spin-orbit evolution under different assumptions about tides, we use the module Planetary Orbital Evolution due to Tides (POET hereafter) (\cite{Penev_2014} for details). POET uses dynamically adjusted eccentricity expansion for the rate of tidal spin-orbit coupling, allowing precise calculations of the evolution even if the eccentricity is high. Furthermore, POET uses a stellar evolutionary model that accounts for evolving stellar structure, spin down of stars driven by magnetic coupling to mass lost via stellar winds, and exchange of angular momentum between the radiative core and the convective envelope.  In order to fully account for  the uncertainties in the observational data for stellar and orbital properties of binary systems, we use detailed Markov Chain Monte Carlo simulations (MCMC) to obtain final constraints on $Q_{\star}^{'}$

The paper is organized as follows: in Section~\ref{sec:2} we describe how the input dataset used for our analysis was constructed. Section~\ref{sec:3} describes the model POET uses for binary evolution with tides. Section~\ref{sec:4} gives the details of our methodology for calculating the likelihood in MCMC. Section~\ref{sec:5} reports the constraints obtained. In Section~\ref{sec:6} we discuss the implication of our results and the assumptions made in our calculations. We conclude in Section~\ref{sec:7}

\section{Data Collection}\label{sec:2}

\subsection{\textbf{Orbital Parameters and Stellar Parameters}}\label{sec:2.1}

We start with the 2165 eclipsing binaries recorded in the second data release of the Kepler Mission \citep{Slawson_2011}. We restrict our sample to binaries with an orbital period of $P_{orb}<45$ days as tidal dissipation effects becomes negligible at large orbital periods. \cite{Kjurkchieva_et_al_17} searched for detached eclipsing binaries with eccentric orbits from \texttt{Kepler} EB catalog \citep{Slawson_2011} and used the code PHOEBE \citep{Pr_a_2005} to generate synthetic light-curves to find best fit estimates for eccentricity and periastron angle. The authors also use the empirical relationship for stellar parameters calculated in \cite{ivanov2010light} to obtain: effective temperature ratio: $T_2/T_1 = (d_2/d_1)^{1/4}$ and mass ratio: $q = (T_2/T_1)^{1.7} = (d_2/d_1)^{0.425}$. We adopted their calculated values for eclipse depths. Since depth measurements using the light-curves from Kepler missions are extremely precise, we ignore the uncertainty in the mass ratio in our analysis. For eccentricities, we used the precision levels reported in their paper as uncertainties, for e>0.1: $\sigma=0.001$; e<0.1: $\sigma=0.01$; and e<0.01 $\sigma=0.1$. Note that \cite{Kjurkchieva_et_al_17} specifically selected eccentric binaries for their analysis. Hence, our dataset is also limited to binaries having eccentric orbits even at low orbital periods.  During the time of this analysis, POET could only handle binary evolution for systems with eccentricities e<0.45, hence we discarded systems with higher eccentricities from the catalog.

Although we would require stellar masses and ages of the binary system to carry out binary evolution, it was difficult to obtain consistently determined mass and age distribution for the eclipsing binaries we selected. Instead, we searched through literature to get distributions for the stellar properties: effective temperature, surface gravity and metallicity of the selected systems and infer mass and age directly from them. \cite{Mathur_2017} published stellar properties of 197,096 \textit{Kepler} targets observed between Quarters 1-17. We cross-matched this catalog with our filtered data set to obtain complete stellar and orbital parameters for our systems of interest. POET handles stellar structure evolution by interpolating among a grid of models generated using Modules for Experiments in Stellar Astrophysics (MESA \cite{Paxton_et_al_11}). Currently this grid is limited to masses: 0.4M< M<1.2M and metallicities: -1.014< Z <0.537. We also take into account these limits and further reduce our dataset.

\subsection{\textbf{Spin Period}}\label{sec:2.2}
Our likelihood calculation requires rotation period measurements for the primary star. \cite{Lurie_2017} performed an extensive study of the light-curves for \textit{Kepler} eclipsing binaries to obtain  rotation periods. The authors identified 816 Kepler eclipsing binary systems with starspot modulations and performed Lomb-Scargle Periodogram \citep{1976Ap&SS..39..447L,1982ApJ...263..835S} analysis to identify multiple peaks of periodic variability in out-of-eclipse light-curves. The multiple peaks in the periodograms come from solar-like differential rotation of the stars. They also used Auto-Correlation function, following the procedure of \cite{McQuillan_2013}, for  validating that the peaks obtained from the periodograms corresponds to periodic variation due to starspots. Following the Lomb-Scargle Periodogram method, they identified two significant peaks (peaks with height >30$\%$ of the highest peak) in their periodograms and classify them into two groups. In each group they then select a subpeak with the largest frequency separation in the neighborhood of the dominant peak. As the authors themselves mentioned, the rotation period corresponding to the highest peak obtained in the first group ($P_{1min}$ in Table 2 of \cite{Lurie_2017}) will be the closest to the equatorial rotation period, we adopt this value as the nominal value of rotation period of the primary star in our analysis. To obtain the uncertainty on the rotation period, we select the rotation period corresponding to the subpeak in the same group ($P_{1max}$ in Table 2 of \cite{Lurie_2017}) and calculate the difference between $P_{1max}$ and $P_{1min}$. After selecting the binaries for which we have the orbital and stellar parameters (Section~\ref{sec:2.1}), we were left with 41 systems.   Table \ref{tab:catalog} shows the full combined catalog used for our analysis.

\section{Binary Evolution using POET}\label{sec:3}

\textbf{P}lanerary \textbf{O}rbital \textbf{E}volution due to \textbf{T}ides (POET) is a module developed by \cite{Penev_2014} to simulate the evolution of binary systems (star-planet and star-star) incorporating both angular momentum changes in stars due to tides and stellar structure changes. As presented in \cite{Penev_2014}, POET only supported circular star-planet systems. With recent developments, it can now also simulate tidal evolution of eccentric exoplanet systems and binary star systems with a very general prescription for the tidal quality factor ($Q_{\star}^{'}$) in the form of a frequency-dependent lag (See Section \ref{sec:6.3.1}). There exists numerous tidal models proposed throughout the literature for evaluating the evolution of binary systems under the influence of tides.

\afterpage{
\begin{landscape}
  \begin{table}
  \centering
  \begin{tabular}{cccccccccccccc}
  \hline
  KIC & $T_{\texttt{eff}}$ & $\sigma_{T_{\texttt{eff}}}$ & [Fe/H] & $\sigma_{[Fe/H]}$ & $P_{\texttt{orb}}$ & $\sigma_{P_{\texttt{orb}}}$ & e & $\sigma_{e}$ & $\log{g}$ & $\sigma_{\log{g}}$ & q & $P_{\texttt{star}}$ & $\sigma_{P_{\texttt{star}}}$ \\
   & (K) & & & & (days) & & & & & & & (days) \\
  \hline
  1026032      & 5951.0             & 160.0                       & -1.06          & 0.30                     & 8.460               & 0.0000230                   & 0.042        & 0.010                  & 4.638              & 0.032  & 0.6630        & 10.778              & 0.384         \\
  3098194      & 5527.0             & 166.0                       & -0.02          & 0.25                    & 30.477             & 0.0001451                   & 0.306        & 0.001                 & 4.315              & 0.205                       & 1.0149       & 26.521              & 6.749           \\
  3323289      & 5427.0             & 161.0                       & -0.10           & 0.30                     & 33.693             & 0.0001642                   & 0.255        & 0.001                 & 4.448              & 0.117                       & 0.8326       & 29.317              & 8.070           \\
  3348093      & 4478.0             & 156.0                       & 0.16           & 0.25                    & 7.964              & 0.0000629                   & 0.124        & 0.001                 & 4.578              & 0.060                        & 0.8000         & 8.185               & 0.272          \\
  4149684      & 6098.0             & 193.0                       & -0.08          & 0.25                    & 4.321              & 0.0000091                   & 0.041       & 0.010                  & 4.384              & 0.090                        & 0.4755       & 4.218               & 0.131          \\
  4352168      & 5282.0             & 159.0                       & -0.70           & 0.30                     & 10.644             & 0.0000319                   & 0.184        & 0.001                 & 4.578              & 0.084                       & 0.5617       & 10.188              & 0.596           \\
  4940201      & 5496.0             & 164.0                       & -0.38          & 0.35                    & 8.817              & 0.0000244                   & 0.046        & 0.010                  & 4.607              & 0.032                       & 0.9269       & 10.296              & 0.753           \\
  5091614      & 6079.0             & 169.0                       & -0.20           & 0.25                    & 21.142             & 0.0000845                   & 0.151        & 0.001                 & 4.495              & 0.054                       & 0.6696       & 20.101              & 2.228           \\
  5347784      & 5605.0             & 168.0                       & -0.04          & 0.30                     & 9.584              & 0.0000272                   & 0.020         & 0.010                  & 3.813              & 0.608                       & 0.9680        & 10.112              & 0.166         \\
  5652260      & 5743.0             & 169.0                       & -0.14          & 0.30                     & 8.929              & 0.0000737                   & 0.022        & 0.010                  & 4.553              & 0.035                       & 0.7898       & 10.953              & 1.166           \\
  5771589      & 6180.0             & 173.0                       & -0.34          & 0.35                    & 10.739             & 0.0000332                   & 0.077        & 0.010                  & 4.007              & 0.292                       & 0.6060        & 14.581              & 0.171          \\
  5986209      & 5559.0             & 183.0                       & 0.04           & 0.25                    & 23.738             & 0.0001259                   & 0.348        & 0.001                 & 4.394              & 0.139                       & 0.5620        & 16.836              & 2.618           \\
  6231401      & 6063.0             & 182.0                       & -0.78          & 0.30                     & 6.092              & 0.0000144                   & 0.028        & 0.010                  & 4.424              & 0.135                       & 0.7660        & 6.055               & 0.105          \\
  6449552      & 5547.0             & 166.0                       & -0.10           & 0.30                     & 20.149             & 0.0000795                   & 0.261        & 0.001                 & 4.447              & 0.098                       & 0.5189       & 18.334              & 3.068           \\
  6468938      & 6047.0             & 163.0                       & -0.32          & 0.30                     & 7.217              & 0.0000181                   & 0.036        & 0.010                  & 4.246              & 0.220                        & 0.9135       & 6.813               & 0.516                        \\
  6579806      & 6113.0             & 164.0                       & -0.30           & 0.30                     & 9.880               & 0.0000288                   & 0.238        & 0.001                 & 4.512              & 0.052                       & 0.4309       & 12.298              & 0.771           \\
  6781535      & 5857.0             & 79.0                        & -0.04          & 0.15                    & 9.122              & 0.0000258                   & 0.252        & 0.001                 & 4.396              & 0.090                        & 0.7695       & 8.031               & 0.218          \\
  6949550      & 5973.0             & 161.0                       & -0.44          & 0.30                     & 7.841              & 0.0000206                   & 0.266        & 0.001                 & 4.562              & 0.037                       & 0.9744       & 6.934               & 0.386           \\
  7846730      & 6271.0             & 199.0                       & -0.10           & 0.25                    & 11.028             & 0.0000332                   & 0.040       & 0.010                  & 4.199              & 0.190                        & 0.8263       & 13.713              & 0.466           \\
  7987749      & 5575.0             & 167.0                       & -0.26          & 0.30                     & 17.031             & 0.0000629                   & 0.142        & 0.001                 & 4.066              & 0.476                       & 0.7188       & 17.545              & 0.449         \\
  8098300      & 6502.0             & 146.0                       & -0.40           & 0.25                    & 4.306              & 0.0000086                   & 0.035        & 0.010                  & 4.44               & 0.052                       & 0.8539       & 4.306               & 0.025           \\
  8296467      & 5497.0             & 166.0                       & -0.28          & 0.30                     & 10.303             & 0.0000306                   & 0.278        & 0.001                 & 4.423              & 0.149                       & 0.8642       & 9.242               & 0.234          \\
  8356054      & 5431.0             & 189.0                       & -0.50           & 0.30                     & 17.081             & 0.0000795                   & 0.219        & 0.001                 & 4.623              & 0.035                       & 0.6979       & 16.708              & 2.344            \\
  8364119      & 5623.0             & 152.0                       & -0.40           & 0.30                     & 7.736              & 0.0000200                   & 0.020         & 0.010                  & 4.596              & 0.036                       & 0.9184       & 8.954               & 0.221          \\
  8374499      & 6099.0             & 164.0                       & -0.28          & 0.30                     & 5.252              & 0.0000116                   & 0.044        & 0.010                  & 4.511              & 0.050                        & 0.4958       & 6.044               & 0.225          \\
  8543278      & 5085.0             & 136.0                       & -0.58          & 0.30                     & 7.549              & 0.0000195                   & 0.018        & 0.010                  & 4.679              & 0.028                       & 0.6828       & 8.603               & 0.505           \\
  8559863      & 5315.0             & 159.0                       & -0.44          & 0.30                     & 22.470              & 0.0000920                   & 0.035        & 0.010                  & 4.636              & 0.030                        & 0.9188       & 38.881              & 4.744                        \\
  8848271      & 5874.0             & 158.0                       & -0.12          & 0.30                     & 9.992              & 0.0000291                   & 0.283        & 0.001                 & 4.338              & 0.153                       & 0.5757       & 13.575              & 0.802           \\
  8938628      & 5815.0             & 157.0                       & -0.62          & 0.30                     & 6.862              & 0.0000170                   & 0.003        & 0.100                   & 4.349              & 0.220                        & 0.8182       & 19.304              & 2.689                        \\
  8984706      & 5895.0             & 175.0                       & -0.32          & 0.30                     & 10.135             & 0.0000300                   & 0.014        & 0.010                  & 4.103              & 0.344                       & 0.9650        & 12.482              & 1.296           \\
  9411943      & 5880.0             & 158.0                       & -0.40           & 0.30                     & 3.844              & 0.0000074                   & 0.048        & 0.010                  & 4.296              & 0.225                       & 0.6987       & 3.828               & 0.050        \\
  9412462      & 5557.0             & 166.0                       & -0.12          & 0.30                     & 10.187             & 0.0000300                   & 0.043        & 0.010                  & 4.448              & 0.108                       & 0.9023       & 9.801               & 0.452           \\
  9468296      & 4961.0             & 151.0                       & -0.14          & 0.30                     & 5.749              & 0.0000132                   & 0.061        & 0.010                  & 4.545              & 0.078                       & 0.8159       & 5.837               & 0.198          \\
  9468384      & 5831.0             & 163.0                       & -0.06          & 0.25                    & 11.083             & 0.0000723                   & 0.067        & 0.010                  & 4.503              & 0.052                       & 0.5594       & 12.991              & 0.377          \\
  9509207      & 5864.0             & 176.0                       & -0.06          & 0.30                     & 14.200               & 0.0000479                   & 0.024       & 0.010                  & 4.139              & 0.299                       & 0.9750        & 23.85               & 0.632           \\
  9896435      & 4848.0             & 173.0                       & 0.06           & 0.25                    & 18.077             & 0.0002024                   & 0.008       & 0.100                   & 4.55               & 0.065                       & 0.5516       & 25.894              & 5.414           \\
  10215422     & 5644.0             & 169.0                       & -0.06          & 0.30                     & 24.847             & 0.0001079                   & 0.290         & 0.001                 & 4.55               & 0.035                       & 0.5022       & 24.83               & 2.224                        \\
  10711551     & 6022.0             & 162.0                       & -0.20           & 0.30                     & 10.692             & 0.0000336                   & 0.053        & 0.010                  & 4.49               & 0.054                       & 0.8791       & 10.692              & 0.315          \\
  10874926     & 6014.0             & 162.0                       & -0.08          & 0.25                    & 11.703             & 0.0000367                   & 0.302        & 0.001                 & 4.494              & 0.052                       & 0.7199       & 14.752              & 2.987          \\
  10992733     & 5492.0             & 180.0                       & 0.20            & 0.20                     & 18.526             & 0.0000710                   & 0.380         & 0.001                 & 4.533              & 0.044                       & 0.8396       & 15.634              & 0.269           \\
  12013615     & 6060.0             & 184.0                       & -0.02          & 0.25                    & 8.203              & 0.0001952                   & 0.011        & 0.010                  & 4.059              & 0.329                       & 0.9018       & 9.361               & 0.203 \\         
  \hline
  \end{tabular}
  \caption{$T_{\texttt{eff}}$: Effective temperature of primary star. 
  [Fe/H]: Metallicity of the primary star. 
  $P_{\texttt{orb}}$:Orbital Period. 
  e: eccentricity.
  $\log{g}$: Surface gravity of primary star. 
  q: Mass ratio. 
  $P_{\texttt{star}}$: Rotation period of primary star.}
  \label{tab:catalog}
  \end{table}  
  \end{landscape}
  
}

\cite{Bolmont_mathis_2016} presented an evolutionary model, accounting for both stellar and orbital changes through time, to study orbital dynamic of close-in planets. The authors simplified the evolution by assuming aligned circular orbits for exoplanet and averaging over the frequency \citep{Ogilvie_13}. To take into account the effects of dynamical tide (enhanced dissipation due to inertial waves for their case), they use a constant time lag model to calculate the dissipation factor. \cite{Benbakoura2019} evaluates an effective tidal quality factor as a sum of quality factors corresponding to equilibrium tides and dynamical tides ($1/Q^{'} = 1/Q^{'}_{eq} + 1/Q^{'}_{dyn}$). The authors in their case again assumes aligned circular orbit and follows the frequency averaged formalism of \cite{Ogilvie_13} to calculate $Q^{'}_{dyn}$. Both of the models are effective in calculating the orbital evolution of close-in planets but are restricted to circular aligned orbits

\cite{Ahuir_strugarek_2021} provides model which takes both tidal and magnetic interactions into account to study evolution of star-planet systems. The authors use numerical model called ESPEM (French acronym for Evolution of Planetary Systems and Magnetism; see \cite{Benbakoura2019}), which assumes a coplanar and circular orbits for the binary system. Similar to POET, this model assumes the star to be divided into two zones: a convective envelope and a radiative core and the authors considers the tidal dissipation  only in the stellar envelope. For equilibrium tide, EPSEM rely on constant value for tidal dissipation parameter throughout the evolution (similar to this work) and for the dynamical tide the authors perform frequency average of the dissipation following the prescription of \cite{Ogilvie_13}, \cite{Mathis_2015}, and \cite{Barker_2020}. This gives the total quality factor as $1/Q^{'} = 1/Q^{'}_{eq} + 1/Q^{'}_{dyn}$. \cite{Attia2021} developed a detailed module, JADE, which calculates evolution of exoplanets which takes into account the photo-evaporation due to the star, the effects of tides on orbital dynamics, presence of a distant pertubator and post-Newtonian relativistic corrections. JADE is relatively detailed model for orbital evolution of exoplanets than POET. In our case, the main focus is the primary star in a binary star system, for which POET is sufficiently adequate.

POET uses a generalised framework for tidal dissipation and thus the tidal quality factor($Q_{\star}^{'}$) to improve upon the already existing models for tidal dissipation including high eccentricities and obliquities. In order to handle eccentric and/or inclined orbits using secular orbital evolution, POET expands the tidal potential each object in the binary experiences in a Fourier series. Each zone of each object then calculates the tidal torque and power of each term in the Fourier series separately, using an effective $Q_\star'$ specific to that zone and term. This allows POET to correctly handle arbitrary frequency/amplitude/spin dependent dissipation, for each zone, making it possible to correctly follow the orbital evolution even for dynamical tide models, including effects like resonance locking or g-mode wave breaking even for significantly eccentric or inclined orbits.

\subsection{Stellar Evolution}
POET uses stellar evolution tracks generated using  MESA \citep{Paxton_et_al_11}  calculated at a grid of masses and metallicities. A distinction is made between the surface convective zone and the radiative core, with the spin of each zone tracked separately. The tracks provide the evolution of  radius, mass and the moment of inertia of each zone separately over time. The grid of tracks is then interpolated to provide values of the stellar quantities of interest (stellar radii, moment of inertia, effective temperature, surface gravity or stellar density) at arbitrary mass, age and metallicity. POET also tracks the exchange of angular momentum between the core and the envelope, assuming the two zones' spins approach solid body rotation on the core-envelope coupling timescale (see Table~\ref{tab:fixed_params}). We use the analysis of \cite{Gallet_Bouvier_15} to fix this timescale as 5Myr.

In addition to the tidal torque, the convective envelope of the star also experiences torque due to magnetic winds. (cf. \cite{Schatzman,Irwin_Bouvier_2010,Gallet_Bouvier_2013}). Combining the efforts of \cite{1987ApJ...318..337S,1988ApJ...333..236K,1996ApJ...462..746B}, POET adopts the following formalism to take these effects into account:
\begin{equation}\label{eq:wind_torque}
\dot{L}_{wind} \equiv -K\omega_{surf}\mathrm{min}(\omega_{surf},\omega_{sat})^2 \left(\dfrac{R_{\star}}{R}\right)^{1/2} \left(\dfrac{M_{\star}}{M}\right)^{-1/2} 
\end{equation}
Here K is the parameterized wind strength, $\omega_{surf}$ is the frequency of stellar surface and $\omega_{sat}$ is the frequency above which the magnetic braking saturates. We fix these values to be constant in our analysis and adopt the wind parameters derived in \cite{Irwin_et_al_07}. There also exist more generalised models for angular momentum changes in star due to stellar winds. \cite{Matt_2015} explores a parametric model which include the dependence of magnetic field strength on stellar surface and global mass outflow rate of the torque experienced by the star due to stellar winds.

Table~\ref{tab:fixed_params} reports the values of all the magnetic braking parameters we assumed to be constant during the evolution. The disk dissipation age is the age at which the initial disk around the primary star dissipates \citep{Irwin_et_al_07,Gallet_Bouvier_15}. The spin frequency of the stellar envelope is held constant until this initial disk dissipates, after which the star evolves according to magnetic torque governed by equation \ref{eq:wind_torque}, combined with the tidal and core-envelope coupling torques.

\begin{table}
  \centering
  \resizebox{\columnwidth}{!}{%
  \begin{tabular}{|c|c|c|}
  \hline
  Disk Dissipation Age             & $t_{disk}$     & 5Myr \\ \hline
  Wind Saturation Frequency        & $\omega_{sat}$ & 2.54 rad/day \\ \hline
  Wind Strength                    & K $\dfrac{\omega^2_{\odot}}{M_{\odot}R^2_{\odot}}$   & $1.26 \times 10^9 Gyr^{-1}$ \\ \hline
  Core-Envelope Coupling Timescale & $t_{coup}$     & 5Myr \\ \hline
  \end{tabular}%
  }
  \caption{Magnetic breaking parameters used in our analysis for stellar evolution}
  \label{tab:fixed_params}
  \end{table}

\subsection{Orbital Evolution}

POET expands upon the prescription described by \cite{Lai_12} to calculate the tidal potential for eccentric orbits. POET extends this formalism using a Taylor expansion in eccentricity, dynamically adjusting the order of the expansion as eccentricity changes during the evolution. This introduces additional Fourier terms in the tidal potential beyond the 7 identified in \cite{Lai_12} for circular orbits. Each tidal term is allowed to have its own prescription for the phase lag between the tidal potential and the distortion in the object it produces. This allows the phase lags to smoothly depend on tidal frequency ($\tilde{\omega}_{m,s}=s\Omega-m\Omega_s$) and spin frequency ($\Omega_s$). Here, $\Omega$ is the orbital frequency.

Following the prescription described in \cite{Lai_12}, consider a binary system of stars with masses M and M$^{'}$. Let the spin angular momentum $\textbf{S}=S\hat{\textbf{S}}$ and orbital angular momentum $\textbf{L}=L\hat{\textbf{L}}$. The tidal potential at a point \textbf{r}  is given by:

\begin{equation}
	U(\textbf{r},t) = \dfrac{GM^{'}}{|\textbf{r}_{M^{'}}|} \left(\dfrac{\textbf{r}\cdot\textbf{r}_{M^{'}}}{|\textbf{r}_{M^{'}}|^2} - \dfrac{|\textbf{r}_{M^{'}}|}{|\textbf{r}-\textbf{r}_{M^{'}}|}\right)
\end{equation}
where $\textbf{r}_{M^{'}}(t)$ is the vector from centers of $M$ to $M^{'}$.

Using a coordinate system where z-axis is along $\hat{S}$ and the y-axis is along $\hat{S}\times\hat{L}$, the tidal potential to the lowest order of ratio of size of the object and semi-major axis in spherical coordinates with $\textbf{r}=(\zeta,\theta,\phi)$, where $\zeta$ is the radial component, $\theta$ is the polar angle and $\phi$ is azimuthal angle is, is given as:
\begin{equation}
	U(\textbf{r},t)= - \sum_{m,m^{'}} U_{m,m^{'}} \zeta^2 Y_{2,m} (\theta,\phi)\dfrac{a^3}{r^3(t)}\exp[-im^{'}\Delta\phi(t)]
\end{equation}

where $\Delta\phi(t)$ is the angle between the line joining from center of M to center of $M^{'}$ and x-axis in the coordinate system where $\hat{z}=\hat{L}$ and $\hat{y}=\hat{S}\times\hat{L}$. Note that $\Delta\phi(t) = \Omega t$ for circular orbits with angular velocity $\Omega$. 

To take into account elliptical orbits, each term in the potential is further expanded as a Fourier series and are given as:

\begin{equation}
	\dfrac{a^3}{r^3(t)}\exp[-im^{'}\Delta\phi(t)] = \sum_s p_{m^{'},s}\exp(-is\Omega t)
\end{equation}

The tidal potential is now a series of spherical harmonic waves, each with its own frequency term. The $p_{m^{'},s}$ allows POET to calculate higher order eccentricity and obliquity terms. For circular orbits $p_{m^{'},s} = \delta_{m^{'},s}$.

Each $(ms)$-component of tidal potential has its own tidal frequency $\tilde{\omega}_{m,s}$ drives the fluid displacement in the star. These perturbations can be defined using Lagrangian displacement $\xi{m,s}(\textbf{r},t)$ and Eulerian density perturbation $\delta\rho_{m,s}(\textbf{r},t)$:

\begin{align}
	\xi_{m,s}(\textbf{r,t}) &= \dfrac{U_{m,s}}{\omega_0^2}\bar{\xi}_{m,s}(\textbf{r})\exp(-is\Omega t +i\Delta_{m,s}) \label{eq:fluid}\\
	\delta\rho(\textbf{r,t}) &= \dfrac{U_{m,s}}{\omega_0^2}\delta\bar{\rho}_{m,s}(\textbf{r})\exp(-is\Omega t +i\Delta_{m,s})  \label{eq:density}
\end{align}

with $\delta\bar{\rho}_{m,s} = -\nabla.(\rho\bar{\xi}_{m,s})$, $\tilde{\omega}_{m,s} = s\Omega-m\Omega_s$ and $\omega_0\equiv\sqrt{GM/R^3}$ is the dynamical frequency of M. .

The torque on the star and energy dissipation rate can then be calculated as:
\begin{align}
	\textbf{T} &= \int d^3x \delta\rho(\textbf{r},t)\textbf{r}\times[-\nabla U^{*}(\textbf{r},t)] \label{eq:torque} \\
	\dot{E} &= \int d^3x \rho(\textbf{r})\dfrac{\partial\xi(\textbf{r},t)}{\partial t} [-\nabla U^{*}(\textbf{r},t)] \label{eq:E}
\end{align}

Although Equations \ref{eq:fluid} and \ref{eq:density}  are inspired by the equilibrium tide model, POET is capable of including the dynamical tide effects with a suitable prescription of $\Delta_{m,s}$. POET converts $Q_{\star}^{'}$ to $\Delta_{m,s}$ using $\Delta_{m,s} = (15/16\pi Q_{\star}^{'})$. POET allows $Q_{\star}^{'}$ to be defined for each zone in the star separately and can have a general dependence on tidal frequency ($\tilde{\omega}_{m,s}$) or spin frequency ($\Omega_s$) (See Section \ref{sec:6.3.1}) allowing for various the dynamical tide models discussed in Section \ref{sec:1}.

The orbital evolution equations are:
\begin{align}
	\dot{a} &= a\dfrac{-\dot{E}}{E} \\
	\dot{e} &= \dfrac{2(\dot{E}L+2E\dot{L})L(M+M^{'})}{G(MM^{'})^3} \\
	\dot{\theta} &= \dfrac{(T_z + \tilde{T}_z)\sin\theta}{L} -  \dfrac{(T_x + \tilde{T}_x)\cos\theta}{L} -  \dfrac{T_x + \mathcal{T}_x}{S} \\
	\dot{\omega} &=  \dfrac{(T_y + \mathcal{T}_y)\cos\theta}{L\sin\theta} +  \dfrac{T_y + \mathcal{T}_y}{S\sin\theta}
\end{align}

For the analysis presented here, we assume all terms have the same modified phase lag (phase lag divided by the love number, $k_2$). We parameterize this phase lag in terms of the modified tidal quality factor ($Q_{\star}^{'} = Q_{\star}/k_2$).

\begin{figure}
  \centering
  \begin{subfigure}{\columnwidth}
    \includegraphics[width=\linewidth]{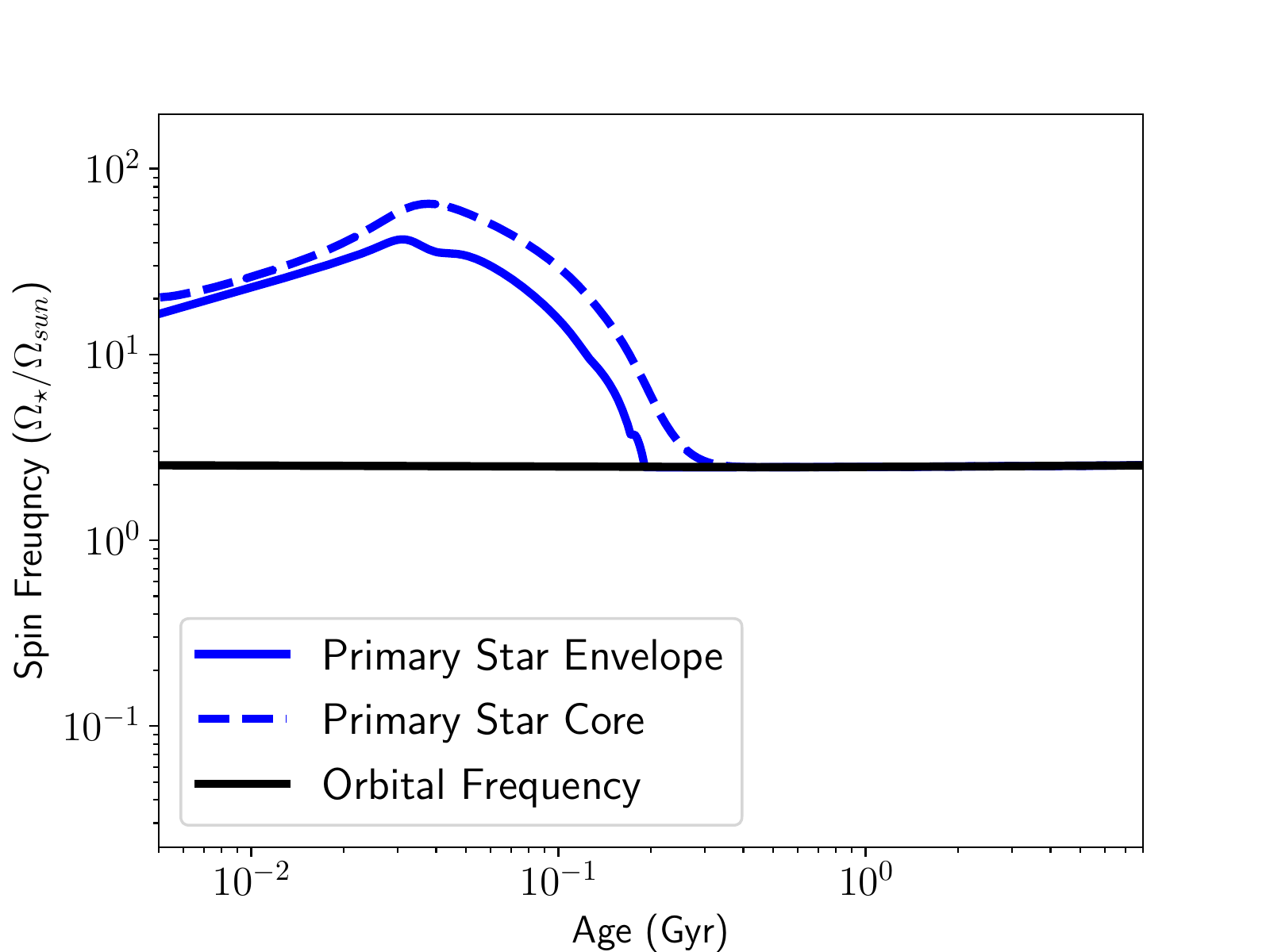}
    \caption{$\log_{10}{Q^{'}_{\star}}=6$ }
    \label{fig:logQ_6}
  \end{subfigure}
  \begin{subfigure}{\columnwidth}
    \includegraphics[width=\linewidth]{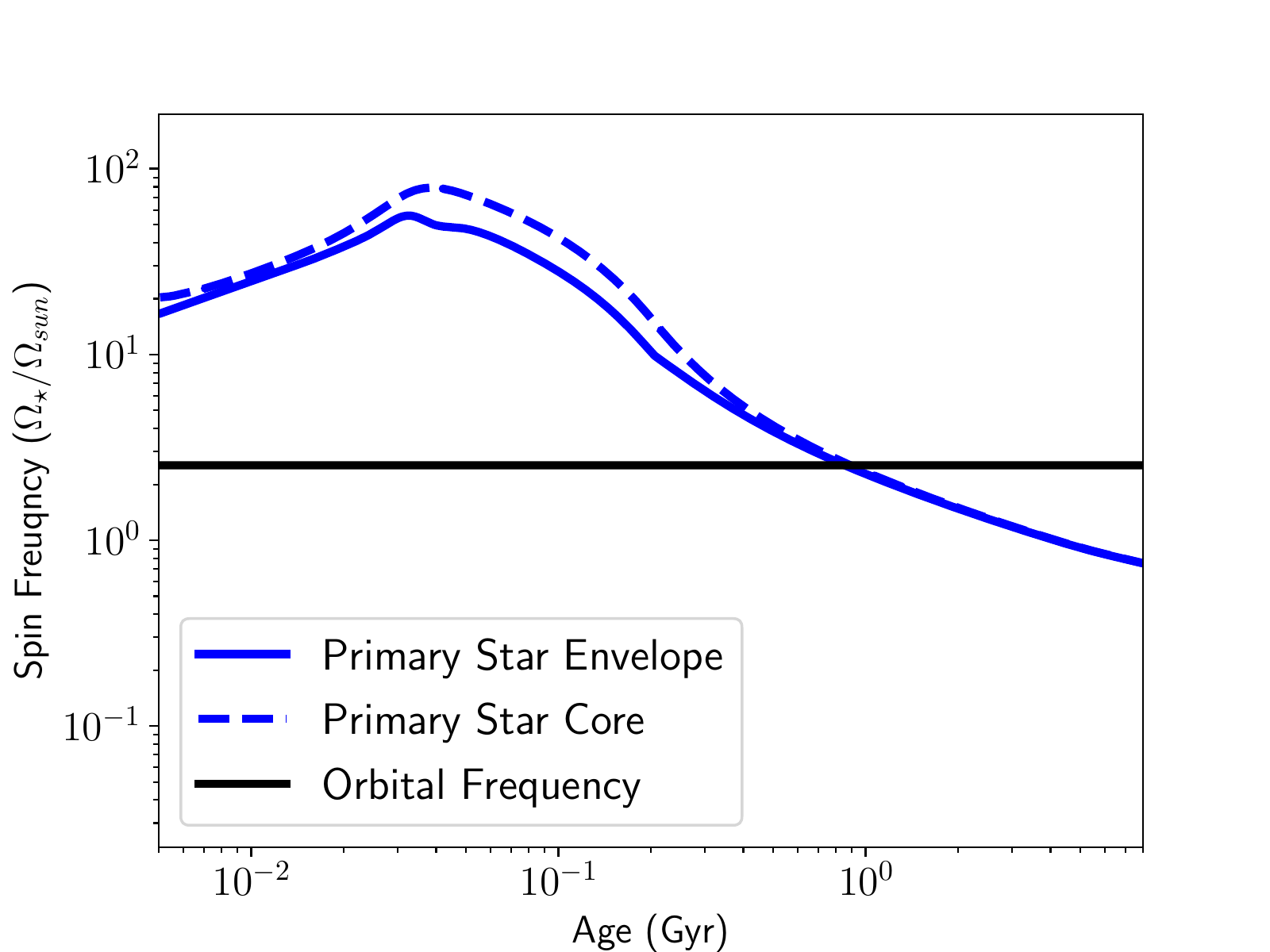}
    \caption{$\log_{10}{Q^{'}_{\star}}=10$ }
    \label{fig:logQ_10}
  \end{subfigure}
  \caption{Evolution of spin frequency of the primary star in a binary system of two solar mass stars under the combined influence of tides and stellar winds at (a)$\log_{10}{Q^{'}_{\star}}=6$ (b)$\log_{10}{Q^{'}_{\star}}=10$. Since the secondary is identical to the primary, it behaves exactly the same. The solid line shows the evolution of the convective envelope and the dotted line shows the evolution of the radiative core. The black line corresponds to the orbital frequency evolution. For $\log_{10}{Q^{'}_{\star}}=6$, the envelope and the core synchronizes with the orbit due to highly efficient tidal dissipation. For $\log_{10}{Q^{'}_{\star}}=10$, since the tides are not as efficient, the spin frequency of the envelope and core evolves similar to isolated star.}
\end{figure}

Fig. \ref{fig:logQ_6} and Fig. \ref{fig:logQ_10} shows the evolution of the spin frequency of one of the star in a binary system of two solar mass stars with $\log_{10}{Q^{'}_{\star}}=6.0$ and $\log_{10}{Q^{'}_{\star}}=10.0$ respectively. Since the secondary is identical to the primary, its evolution is also the same as shown. During the initial stages, the stellar spin is dominated by the changes in stellar structure. The initial rise in spin frequency is due to the shrinking of stellar radius and hence decreasing moment of inertia as the star approaches the main sequence.  On the main sequence the shrinking comes to a halt and the spin of the convective zone of the star evolve on longer timescales of the two torques, due to tides and stellar winds . This leads to the spinning down of the star, bringing spin frequency closer to the orbital frequency. As the core-envelope coupling timescale is assumed to be several Myrs, the core of the star also experiences the same spin down. If tidal dissipation is highly efficient ($\log_{10}{Q^{'}_{\star}}=6$ in Fig. \ref{fig:logQ_6}), the star will be locked into a spin-orbit synchronous spin. On the other hand, if tidal dissipation is low ($\log_{10}{Q^{'}_{\star}}=10$ in Fig. \ref{fig:logQ_10}), the star evolves as isolated main sequence stars. Note that, in the case of high tidal dissipation, both stars and the orbit exchange angular momentum. Hence, if the stars spin down, the orbital frequency decreases as a result. However, in the case of a binary star system, due to the high mass of the secondary companion (in this case, a solar mass star), the change in orbital frequency is small as compared to the changes in spin frequency.

\section{Methodology}\label{sec:4}

We use Markov Chain Monte Carlo (MCMC) simulations for systems listed in Table~\ref{tab:catalog} to take into account the impact of observational uncertainties on the inferred value of $Q_{\star}^{'}$. The likelihood is calculated using the observed rotation period of the primary star and the corresponding 1-$\sigma$ error listed in columns 13 and 14 of Table~\ref{tab:catalog}. Although we allow dissipation in both stars, the value of $Q^{'}_{\star}$ is assumed to be constant in time and the same for both the stars. Before running the MCMC, we find a maximum likelihood estimate for $\log_{10}{Q^{'}_{\star}}$  by using the nominal values of each parameter and solving for $\log_{10}{Q^{'}_{\star}}$ value which reproduces the nominal value of the spin period. The main steps for calculating the likelihood are summarized as follows:

\begin{enumerate}
    \item Using effective temperature and surface gravity, infer samples for mass, age and metallicity. 
    \item Draw random samples for the sampling parameters described in Section~\ref{sec:4.1}.
    \item Draw a random value of $\log_{10}{Q^{'}_{\star}}$ from a uniform distribution.
    \item Find the initial orbital period and initial eccentricity, which when evolved to the sampled age of the system matches the orbital period and eccentricity sampled in step (ii). The initial conditions thus found also gives the evolution that predicts a particular value for the present day spin of the star.  
    \item Get the likelihood assigned to the sampled parameters from a Gaussian distribution based on the observed orbital period and its uncertainties, evaluated at the predicted stellar spin in step (iv)
\end{enumerate}

\subsection{Sampling Parameters}\label{sec:4.1}
We sample from a set of 7 parameters. 
\begin{itemize}
    \item \textbf{Orbital Period ($P_{orb}$) and Eccentricity ($e$):} These are the present day orbital period and eccentricity reported in Table~\ref{tab:catalog}. We assume Gaussian priors for both the parameters based on the observational error bars. 
    \item \textbf{Initial Disk Frequency ($W_{disk}$):} POET calculates the evolution of the stars in binary system, under the influence of tides, after the disk dissipation age, fixed at 5Myr(as mentioned in Section \ref{sec:2.2}). Before this age, both the stars evolve as isolated stars with an initial disk around them. Before the disk dissipates and the binary forms, the spin angular frequency of the convective zone of the star is held equal to the disk frequency, while the spin of the core is allowed to evolve under moment of inertia changes and core-envelope coupling. The final rotation period will not be influenced by this initial period if the age of the binary system is large enough \citep{Gallet_Bouvier_15}. We assume uniform prior for the disk frequency:
$$\Pi(W_{disk}) = \textit{U}(1.4 \ rad/day,4.4 \ rad/day)$$
where $W_{disk}$ is the disk frequency, and the limits are selected from \cite{Gallet_Bouvier_15}. 

    \item \textbf{Mass, Age, and Metallicity:} We assume that both the stars in the binary system were formed at the same time from the same molecular cloud. Hence, we use the distribution of metallicity of the primary star reported in Table~\ref{tab:catalog} and assume both stars have exactly the same metallicity. Furthermore, the values of the mass of the primary star and the age of the system were inferred from other parameters as explained in Section~\ref{sec:4.2}.
    
    \item \textbf{log$_{10}$Q$^{'}_{\star}$}: We select uniform priors on $\log_{10}{Q^{'}_{\star}}$
$$\Pi(\log_{10}{Q^{'}_{\star}}) = \textit{U}(5.0,12.0)$$
\end{itemize}

\subsection{Sampling Mass,Age and Metallicity}\label{sec:4.2}
The mass of the primary star and age of the binary system are inferred from  primary star's effective temperature ($T_{\texttt{eff}}$) and surface gravity ($\log_{10}{g}$). We run separate MCMC simulations for each system to directly obtain marginalized samples for mass, age and metallicity. The priors on these parameters are:
$$\Pi(M) = \textit{U}(0.4M_{\odot},1.2M_{\odot})$$
$$\Pi(T) = \textit{U}(0.001 K,10 K)$$
$$\Pi([Fe/H]) = \mathcal{N}([Fe/H]_0,\sigma) \times \textit{U}(-1.014,0.537)$$
where $[Fe/H]_0$ and $\sigma$ comes from the observations. 
The likelihood computation is as follows:
\begin{enumerate}
    \item Draw random sample directly from the priors for mass, age and metallicity
    \item Compute $T_{\texttt{eff}}$ and $\log_{10}{g}$ using the interpolated grids of isochrones obtained from MESA.
    \item Find the likelihood by assuming a Gaussian distributions for both $T_{\texttt{eff}}$ and $\log_{10}{g}$:
    $$L = N(t|T_{\texttt{eff}},\sigma_{T_{\texttt{eff}}}) \times N(g|\log_{10}{g},\sigma_{\log_{10}{g}})$$
    where the nominal values and 1-$\sigma$ error for both parameters are listed in Table~\ref{tab:catalog}.
\end{enumerate}
The samples thus obtained are now used for the main MCMC. Appendix \ref{sec:app} discusses the full details behind the sampling process and modifications to MCMC made for faster convergence.

\subsection{Testing MCMC Convergence using Geweke's Convergence diagnostic}
    \cite{Geweke_1991} proposed a convergence diagnostic by comparing the mean and variance of different segments, taken from the start and the end of the single MCMC chain. The test statistic, z-score, is given as:
    \begin{equation}
      z = \dfrac{\bar{\theta_i} - \bar{\theta}_f}{\sqrt{Var(\theta_i) + Var(\theta_f)}} 
    \end{equation}
    where \textit{i} is a segment at the beginning of the chain and \textit{f} is a segment at the end of the chain. The final segment of the chain can be further divided into multiple parts. 
    If the z-score for all the parts are similar, we can conclude that the chain is well-mixed and converged.
    
    For our case, since we run multiple chains for each system, we perform this test after combining all the chains together.
    Furthermore, given a set of parameters, the computational time for each iteration in MCMC can hugely vary. This results in chains of unequal length. In order to maintain consistency while selecting starting and final segments, we divide each chain into segments before combining them.
    We select the first 10$\%$ of each chain as the starting segment and the last 50$\%$ as the final segment. The final segments of each chain are divided into 20 parts. Fig. \ref{fig:z_test} shows the z-scores for all the systems. We mark 2-$\sigma$ values from zero as our criteria for convergence.
    
\afterpage{
\begin{landscape}
      \begin{figure}
      \centering
      \includegraphics[width=\linewidth]{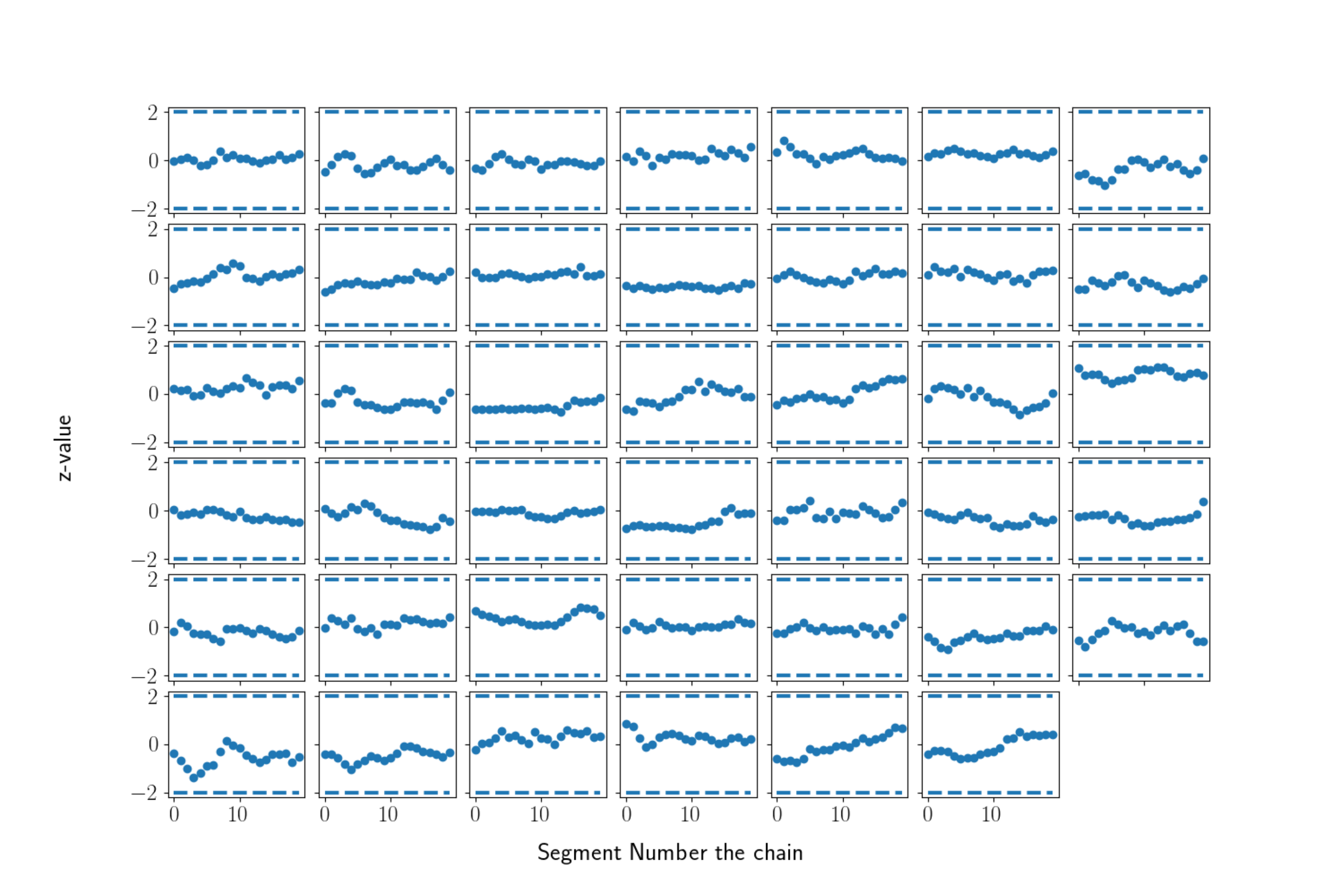}
      \caption{Geweke convergence test for convergence of MCMC chains for all binary systems. Each plot shows z-value calculated by breaking the last 50$\%$ of $\log_{10}{Q_{*}^{'}}$ chain into 20 parts. The  2-$\sigma$ values from zero are taken as our criteria for convergence.}
      \label{fig:z_test}
    \end{figure}
\end{landscape}
\begin{landscape}
\begin{figure}
  \centering
  \includegraphics[width=\linewidth]{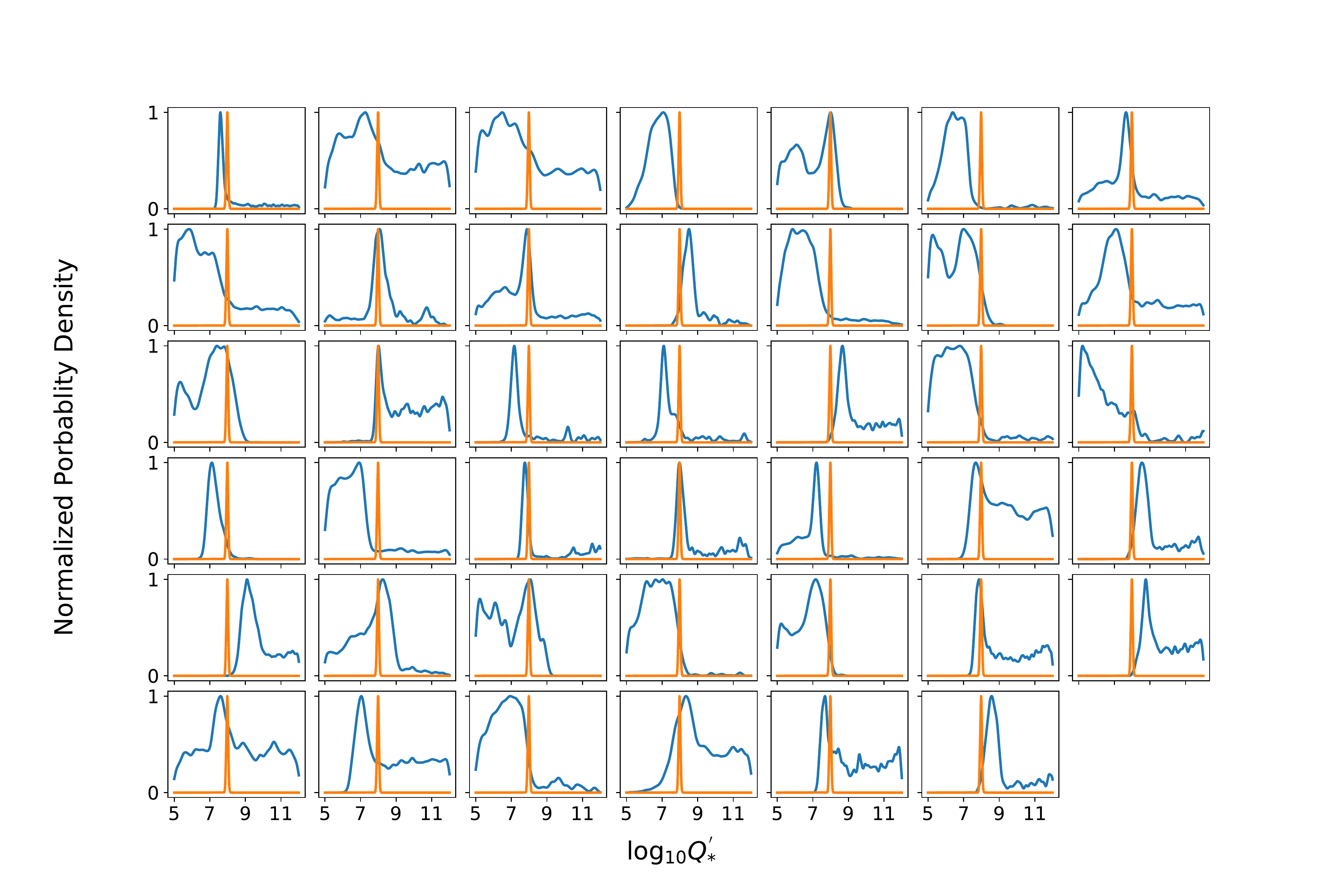}
  \caption{$\log_{10}{Q^{'}_{\star}}$ constraints obtained for the binary systems listed in Table~\ref{tab:catalog}. Each curve shows normalized probability distribution of $\log_{10}{Q^{'}_{\star}}$ where $\log_{10}{Q^{'}_{\star}} \in [5,12]$.  The blue curves are the individual probability density functions constructed using a Gaussian kernel density estimator on the $\log_{10}{Q^{'}_{\star}}$ samples obtained from MCMC. The orange curve is a joint distribution obtained by multiplying all of the individual curves together. The joint distribution can provide common $\log_{10}{Q^{'}_{\star}}$ constraints, valid for all the binary systems in our analysis. This can be confirmed by quantifying the overlap between these two curves for each system using the expectation of p-values calculated over the whole $\log_{10}{Q^{'}_{\star}}$ range}
  \label{fig:pdf}
\end{figure}

\end{landscape}
}

\section{Results}\label{sec:5}

\subsection{Constraints Obtained for Tidal Quality Factor}\label{sec:5.2}

The probability density function for $\log_{10}{Q^{'}_{\star}}$ is calculated using a Gaussian kernel density estimator:
\begin{equation}\label{eq:pdf}
    P_S(q) = \dfrac{1}{nh}\sum_{i=1}^{n}K(\dfrac{q-q_i}{h})
\end{equation}
where $q_i$ are the posterior samples, n is total number of samples for each system, h is kernel bandwidth given as: $h=3.5n^{-1/3}$ and $K(x)=(1/\sqrt{2\pi})\exp{-x^2/2}$. 
Most of the systems in our analysis produced a one-sided limit for $\log_{10}{Q^{'}_{\star}}$. In order to find an overlapping range of $\log_{10}{Q^{'}_{\star}}$ values for all the systems we calculate a combined joint distribution for $\log_{10}{Q^{'}_{\star}}$ by multiplying the individual probability density functions of each systems:
\begin{equation}\label{eq:jdf}
    J(q) = \Pi_{i=1}^{41}P_i(q)
\end{equation}

The functions defined by equations \ref{eq:pdf} and \ref{eq:jdf} are shown in Fig. \ref{fig:pdf}. We find $\log_{10}{Q^{'}_c}=7.818\pm0.035$ from this joint distribution.

\section{Discussion}\label{sec:6}

\subsection{Common constraints obtained from joint distribution}
If our tidal dissipation and orbital evolution models are valid, the combined $\log_{10}{Q^{'}_c}$ value reported in Section~\ref{sec:5.2} should fall within the $\log_{10}{Q^{'}_{\star}}$ posteriors of all the binary systems. In order to confirm this, we can quantify the overlap between the individual probability distribution functions and the combined distribution of $\log_{10}{Q^{'}_c}$  using the following procedure:
\begin{enumerate}
  \item First, calculate the p-value for each system from the distribution $P_s$ (Equation \ref{eq:pdf}) at a certain $q=\log_{10}{Q^{'}_{\star}}$. This gives the probability of finding q at the tail end of the distribution given in Equation \ref{eq:pdf}:
  \begin{equation}\label{eq:p_value}
    \mathrm{p_s(q)} = 2\mathrm{min(CDF_s(q),1-CDF_s(q))}
  \end{equation}
  \item Next, calculate the expectation of the p-value calculated in Equation \ref{eq:p_value}, assuming q is drawn from combined distribution (J). This is related to the probability of the individual distribution $P_s$ and $J$ probing the same region of $\log_{10}{Q^{'}_{\star}}$-parameter space:
  \begin{equation}\label{eq:expected}
    \mathrm{E[p(q)]_s} = \int_{5}^{12} \mathrm{p_s(q)J(q)}
  \end{equation}
\end{enumerate} 

\begin{figure}
  \centering
  \includegraphics[width=\linewidth]{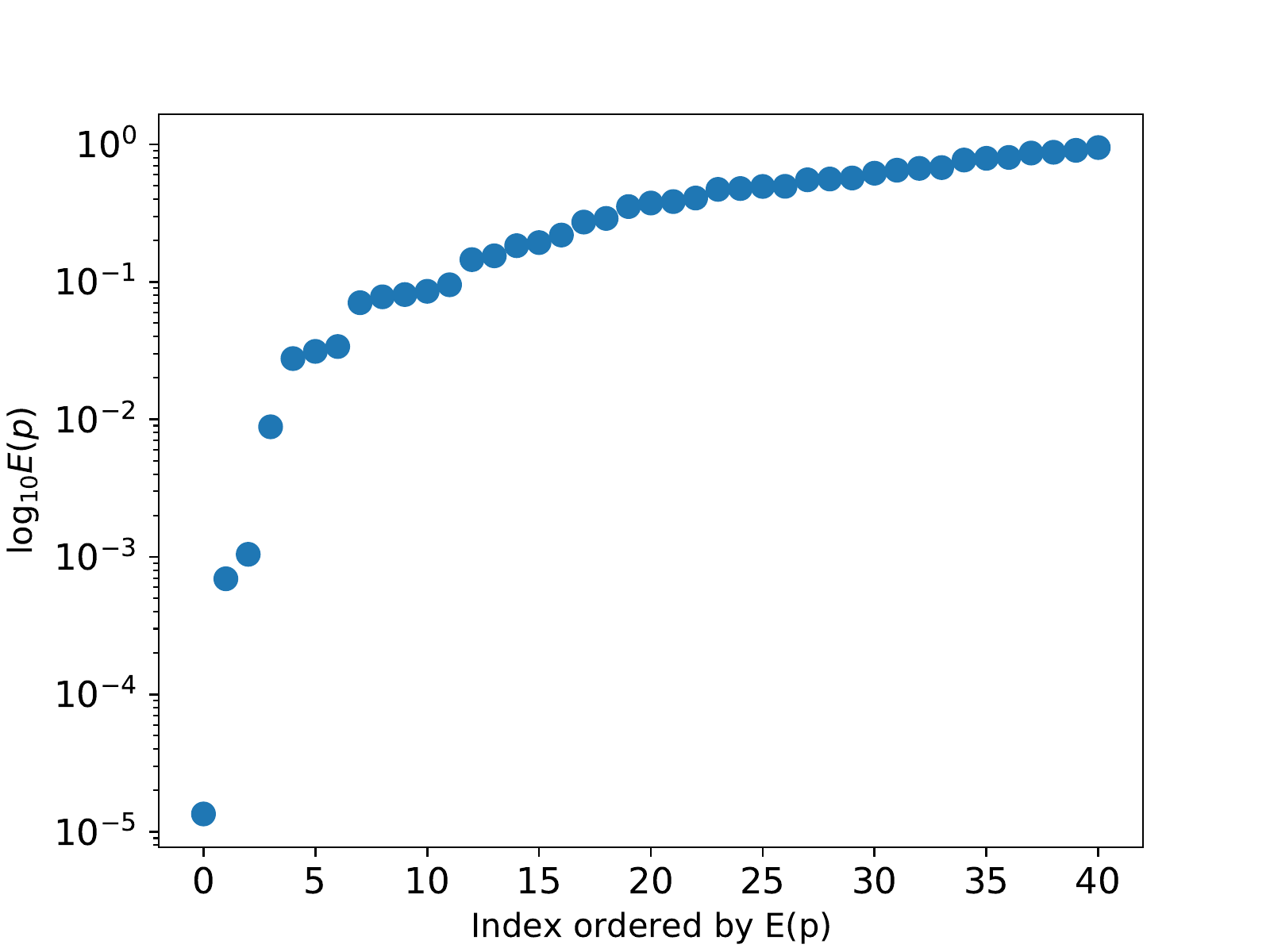}
  \caption{Expected p-values for 41 binary systems used in the analysis. The $\log_{10}{E(p)}$ values are calculated using Equation \ref{eq:expected}. The values quantifies the overlap between individual probability distribution (Equation \ref{eq:pdf}) and the common joint distribution (Equation \ref{eq:jdf}) }
  \label{fig:E_p}
\end{figure}

Although equation \ref{eq:expected} has integration limits from $\log_{10}{Q^{'}_{\star}}=5.0$ to $\log_{10}{Q^{'}_{\star}}=12.0$, the 1-$\sigma$ error for J(q) is so small ($=0.035$) that we can think of J(q) as a delta function at $\log_{10}{Q^{'}_c}=7.818$. Under this assumption Equation \ref{eq:expected} then calculates the standard p-value for each individual distribution. With this context we can interpret values reported in Fig. \ref{fig:E_p} as the probability of a value $\log_{10}{Q^{'}_c}=7.818$ belonging to the probability distribution (\ref{eq:pdf}) of each system we selected. In other words, the p-values tells us whether $\log_{10}{Q^{'}_c}=7.818$ is a good enough value for $\log_{10}{Q^{'}_{\star}}$ such that it can reproduce  the distribution of rotation period for each system.  

Out of the 41 systems, 30 systems have p-values well above 10$\%$. There are three systems which have extremely low p-values (\textless $1\%$): \texttt{KIC 8938628} $\simeq 10^{-5}$, \texttt{KIC 3348093} $\simeq 7\times10^{-4}$ and \texttt{KIC 9509207} $\simeq 10^{-3}$. We flag theses systems as outliers in our analysis. We inspect if these outliers are present because of the enhancement in tidal dissipation due to inertial waves \citep{Ogilvie_2007}. If the tidal frequency is less than twice the spin frequency of the star, i.e $2(\omega_{orb}-\omega_{\star})<2\omega_{\star}$, the system falls within the inertial mode enhancement region. All of our systems fall within the regime except one of our outlier: \texttt{KIC 8938628}

\subsection{Upper and Lower limits on $\log_{10}{Q^{'}_{\star}}$}
From the results in Fig. \ref{fig:pdf}, we see that most systems either have an upper limit or a lower limit on $\log_{10}{Q^{'}_{\star}}$. To explain this, we categorize the systems into synchronized and non-synchronized systems based on the observational data from Table \ref{tab:catalog}. Systems for which the orbital frequency ($\omega_{orb}$) is within the 1-$\sigma$ error of the primary star's spin frequency ($\omega_{star}$) are labeled as synchronized systems. Out of 41 systems, 16 are synchronized and 25 are non-synchronized.

\begin{figure}
  \centering
  \includegraphics[width=\linewidth]{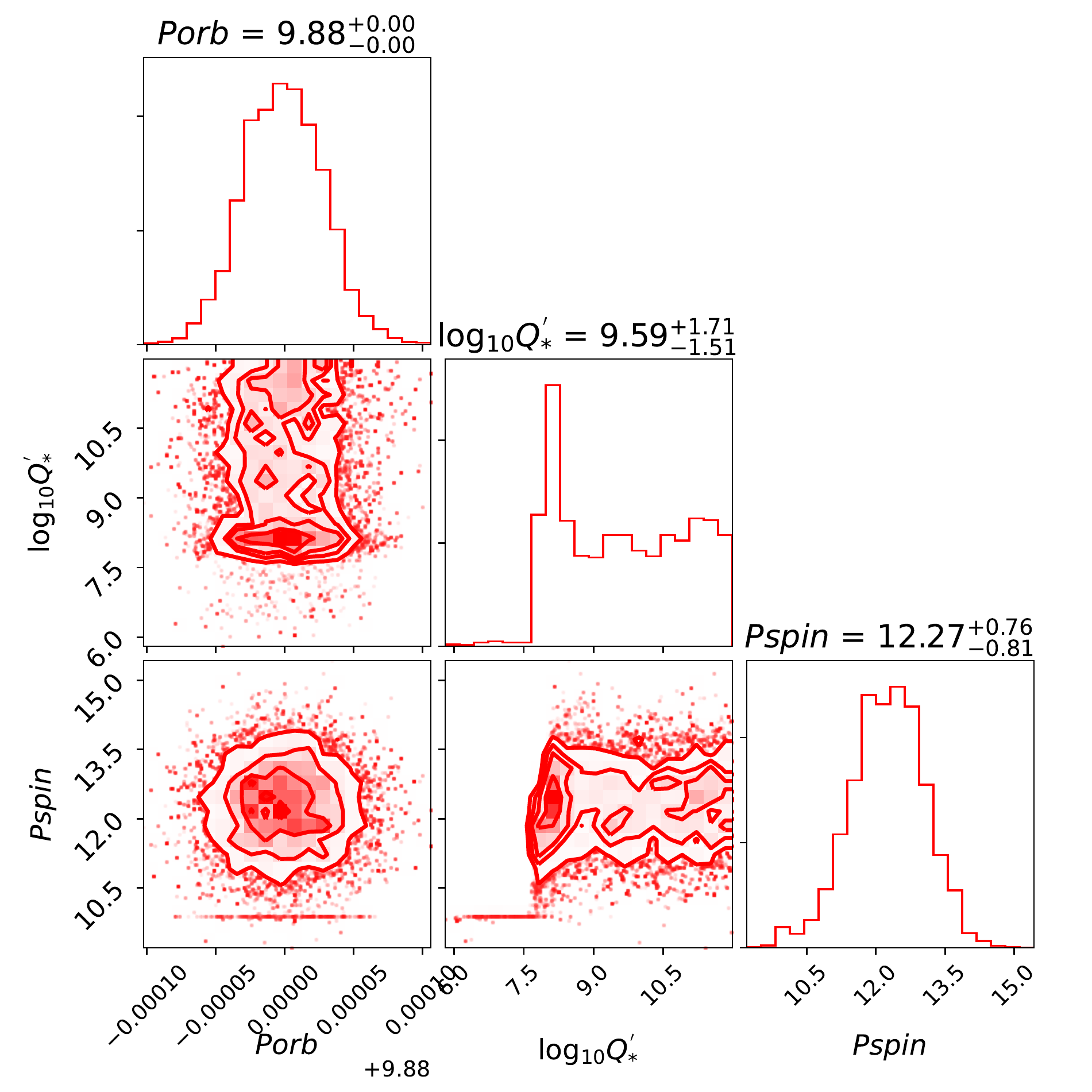}
  \caption{Contour plots for orbital period ($P_{orb}$), $\log_{10}{Q^{'}_{\star}}$ and spin period ($P_{spin}$) for \texttt{KIC 6579806}. At low values of $\log_{10}{Q^{'}_{\star}}$ (high tidal dissipation), the spin period of star synchronized with the orbital period. This prevents non-syncrhonized system to reproduce spin period at low values putting a lower limit to $\log_{10}{Q^{'}_{\star}}$}
  \label{fig:non_sync}
\end{figure}

In the case of non-synchronized systems, small values of $\log_{10}{Q^{'}_{\star}}$ (large tidal dissipation) can be rejected because they would cause the spin of the primary star to synchronize to the orbit. This is shown for \texttt{KIC 6579806} in Fig. \ref{fig:non_sync}. Since the dissipation does not need to be very efficient to reproduce the spin period, there is no upper limit imposed on $\log_{10}{Q^{'}_{\star}}$.

For the systems which are synchronized, an upper limit on $\log_{10}{Q^{'}_{\star}}$ is imposed because at at higher values of $\log_{10}{Q^{'}_{\star}}$ dissipation would not be sufficient to synchronize the primary star with the orbit. There are also few non-synchronized systems in our analysis that shows both a lower limit and an upper limit on $\log_{10}{Q^{'}_{\star}}$. As explained earlier, the lower limit is imposed because of the synchronization of the primary star with the orbit. On the other hand, a possible explanation for the upper limit is that even though the primary star is not synchronized with the orbit, its spin period is still very different from the spin expected of isolated stars with similar properties, ruling out the high $\log_{10}{Q^{'}_{\star}}$ (low dissipation) regime.

\subsection{Caveats}

\subsubsection{Frequency Dependent $\log_{10}{Q^{'}_{\star}}$}\label{sec:6.3.1}

We have assumed a constant $\log_{10}{Q^{'}_{\star}}$ for both stars in our analysis. For non-synchronized circular systems, it is reasonable to assume that the tidal frequency term, $\omega_{tide} = 2|(\omega_{star} - \omega_{orbit})|$, is the most dominant term for tidal dissipation in the star. On the other hand, for non-synchronized eccentric systems, it is difficult to say which is the dominating frequency term. 

In order to understand the dominant dissipation mechanism in the stars we will require a detailed exploration of frequency dependence of $\log_{10}{Q^{'}_{\star}}$. As mentioned in Section \ref{sec:1}. $Q^{'}_{\star}$ can have different power-law dependence depending upon the nature of tidal dissipation. For future analysis we can propose a power-law or a breaking power-law which depends on the tidal frequency. The simplest formulation of frequency dependent $\log_{10}{Q^{'}_{\star}}$ is:
\begin{equation} \label{freqlogQ}
Q^{'}_{\star}= Q_{\star}^{'0}max\left(1, \left(\frac{\omega_{break}}{\omega_{tide}}\right)^{\alpha}\right)
\end{equation}
Here $\omega_{break}$ is a break frequency introduced such that $\log_{10}{Q^{'}_{\star}}$ assumes a constant value below or above it depending upon the sign of $\alpha$. Now, instead of sampling from $\log_{10}{Q^{'}_{\star}}$, the MCMC samples from $\log{Q_0}$, $\alpha$ and $\omega_0$. A more complex formulation would be to assume a broken power-law with different indices and breaks.

POET can also allow a user-defined function for $Q_{\star}^{'}$ with the dependence on wave amplitude, zone boundaries, tidal frequency and spin frequency to deal with more complicated models like resonance locking in case of dynamical tides which enhances tidal dissipation for a dense set of frequencies with sharp peaks in the  frequency spectrum \citep{witte1999,burkart2014,fuller2016}, but we are currently unsure of the numerical stability of the orbital evolution. We reserve the frequency dependent analysis of $Q_{\star}^{'}$ for future work.

\subsubsection{Stalling Spin-Down of K Dwarfs}
Although we account for the differential core-envelope coupling for the angular momentum evolution in our stellar model, we do not account for the stalled spin-down for K stars \citep{Curtis_2019}. The authors expanded the rotation periods of single stars in the open cluster NGC 6811 from \cite{Meibom_2006} to include lower mass stars (up to 0.6 solar masses). They adopt a gyrochronolgy model constructed by \cite{Douglas_2019} for young open cluster Praesepe (670 Myr) and project their fit from 670 Myr to 1 Gyr for NGC 6811. As shown in fig. 5 of \cite{Curtis_2019}, by plotting the color-period distribution of the stars in NGC 6811 over the two models: Praesepe (670 Myr) and NGC 6811 (1 Gyr) it can be observed that while stars with $T_{eff}>5400 K$ have all spun down relative to Praesepe, stars with cooler temperature have not spun down enough and still have rotation period lying on the 670 Myr Praesepe. However, this stalled spin-down disappears for stars with $T_{eff}<4800K$. This is interpreted as  stalling of spin down in lower mass stars (K dwarfs), where the rotation period of the low mass stars stops spinning down and continues after a pause. In our stellar evolution model we do not account for this stalling, which may result in inaccurate calculation of the rotation period of primary star when MCMC samples an age where this stalling happens. Of the 41 systems we analyzed, only 5 systems, \texttt{KIC4352168, KIC8543278, KIC8559863, KIC9468296 and KIC9896435}, have nominal temperatures in range 4800K-5400K which could exhibit such behaviour.

\subsubsection{Dependence of Tidal Quality Factor on stellar properties}

The tidal dissipation is also sensitive to the stellar properties such as mass, age, radius, temperature and metallicity. \cite{Ahuir2021} formulates the dissipation in stellar radiative zones through internal gravity waves. Their study expands through the PMS to the RGB phases of the F-,G- and K-type stars. \cite{Mathis_2015} evaluated the dependence of mass, age and rotation of the star using a simplified two-layer model \citep{Ogilvie_13} to compute frequency-averaged tidal dissipation in the convective envelope of low mass stars (from M to F type stars). The authors showed that, for a fixed angular velocity of the star, during the Pre-Main Sequence phase, the dissipation increases with age and reaches a maximum value. During the Main Sequence phase of the stars, the dissipation decreases and achieves a constant value for the rest of the evolution. \cite{gallet_2017}  used the frequency averaged formalism of \cite{Ogilvie_13} to generalise the work of \cite{Mathis_2015} to include Red Giant Branch of stellar evolution. The authors uses grid of stellar evolution models of rotating stars with initial masses between 0.3 - 1.4 $M_{\odot}$ to evaluate evolution of star from Pre-Main Sequence(PMS) to Red Giant Branch(RGB) phase, to study the effects of tidal dissipation due to dynamical tides in the stellar convective envelopes combined with stellar evolution dude to stuctural changes. Similar to POET, the authors also take into account the star-disc interactions during early-PMS phase and the stellar rotation is held a constant value upto a certain timescale. Following \cite{Matt_2015} prescription, they assume convective region to be in solid body rotation and the magnetic braking is extended from PMS to RGB phase. \cite{Bolmont_mathis_2017} demonstrated the dependence of metallicity on tidal dissipation and \cite{Bolmont_mathis_2016} investigated the dependence of stellar rotation on tidal dissipation.  \cite{Mathis_2015} showed during the pre-main sequence of the stellar evolution, low-mas stars have increased dissipation for a fixed angular velocity.  \cite{Barker_2020}  computed the dependence of the tidal quality factor on stellar mass, age, rotation, tidal frequency and amplitude, combining the effects of inertial modes, gravity modes and turbulent friction. The authors conclude that the depending on the stellar properties, $\log_{10}{Q^{'}_{\star}}$ varies significantly depending on the stellar mass and age. For our analysis we assumed a constant value of $\log_{10}{Q^{'}_{\star}}$ for all the stars throughout the evolution. For the binary systems in our sample, we are only sensitive to the dissipation in the primary star. It can be seen from Table\ref{tab:catalog}, the spread between effective temperature and metallicity of the primary star is not huge. This means the variability in mass and age derived from these parameters (Section \ref{sec:4.2} )
for our analysis is not diverse enough to explore the dependence of $Q_{\star}^{'}$ on stellar parameters. We reserve this analysis for future work once we have a larger sample of binary stars.

\def\beck{\mbox{\cite{Beck2018}}}

\section{Conclusion}\label{sec:7}
We analyzed 41 eccentric low-mass eclipsing binary systems to find constraints on tidal quality factor ($Q_{\star}^{'}$). Our main objective was to use the observed rotation period of the primary star to derive tight constraints on $\log_{10}{Q^{'}_{\star}}$. We used Markov chain Monte Carlo simulations to account for the uncertainties in the observed stellar and orbital parameters available in the literature. By using the module POET(\cite{Penev_2014}), we relaxed numerous assumptions usually made while calculating the evolution of binary systems under the influence of tides. 

Combining the individual constraints from all systems, we constructed a joint constraint, finding a common $Q_{\star}^{'}$ values viable for almost all the binary systems reported in Table~\ref{tab:catalog}. We report $\log_{10}{Q^{'}_{\star}}=7.818\pm0.035$ should be a valid distribution for $Q_{\star}^{'}$ that can reproduce the rotation period distribution of primary star for all system. To quantify this, we calculate expected p-values for each system using the join distribution.  These values are reported in Fig. \ref{fig:E_p}. We find the p-values are reasonable for all systems except \texttt{KIC 8938628}, \texttt{KIC 3348093} and \texttt{KIC 9509207}. Some theoretical models suggest that the tidal dissipation may be much larger if the tidal frequency is in the range where interaction with inertial waves may occur \citep{Ogilvie_2007}. All systems in our sample, except \texttt{KIC 8938628} fall in that range. 

We assumed $Q_\star^{'}$ to be constant in time in our binary evolution model. However, for more detailed analysis this limitation can be relaxed by assuming a frequency-dependent $Q^{'}_\star$ in the form of a power-law or a broken power-law in tidal frequency. Our stellar evolutionary model did not account for the stalling of spin-down in low-mass stars (K-dwarfs) \citep{Curtis_2019}. More accurate measurement of tidal dissipation can be obtained by incorporating the recent development in the understanding of stellar spin-down. POET is also restricted to main sequence stars in binary systems. \beck{} investigate effects of tides on stars on red giant branch in case of both equilibrium and dynamical tide models.

\section*{Acknowledgements}

This research was supported by NASA ATP grant 80NSSC18K1009. 
 The authors acknowledge the Texas Advanced Computing Center (TACC) at The University of Texas at Austin for providing HPC resources that have contributed to the research results reported within this paper. \url{http://www.tacc.utexas.edu}. The authors also acknowledge the use of the \textit{ganymede} HPC cluster at the University of Texas at Dallas.

\appendix
\section{Adjustments in Sampling Process}\label{sec:app}

\subsection{Sampling from Discrete Distribution}\label{sec:8.1}
One caveat of using MCMC is that the proposal function with a user-defined step size could prolong the time taken for the starting distribution to converge to the target distribution. The proposal function for the orbital period, eccentricity, initial disk period and $\log_{10}{Q^{'}_{\star}}$ was assumed to be a Gaussian distribution with a mean equal to the last proposed value and standard deviation equal to the step size. For the samples obtained in Section~\ref{sec:4}, for most of the systems, the distribution in age is almost uniform and extends over the respective prior distribution. To avoid sampling uniformly from the samples of mass, age, and metallicity, and make sure that at each iteration the new proposed samples remain in the vicinity of previously accepted samples, we decided to impose a Gaussian step function for these parameters too. This ultimately leads to an increase in the rate of acceptance and eventually a faster convergence for the MCMC chains.

Since for mass, age and metallicity we now have a set of discrete samples obtained from a different MCMC run (see Section~\ref{sec:4}), we can distinguish between the two types of parameters by considering a discrete set of parameters $\phi$: (mass, age, metallicity) and a continuous set of parameters $\theta$: (orbital period, eccentricity, initial disk period, $\log_{10}{Q^{'}_{\star}}$).

Let $\phi_k$ and $m_k$ be the current parameter set at an iteration N in MCMC and its respective multiplicity in the chain of n samples. Here, multiplicity is defined as the number of times each sample point is repeated (because of rejected MCMC proposals) such that the ratio $\left(\dfrac{m_k}{ \sum_{i=1}^n m_N }\right)$ is the probability of sampling a point $\phi_k$  from the distribution. Let $\sigma_{\phi}$ be the proposed step size. The modification to the proposal function for $\phi_k$ parameters is as follows:

\begin{enumerate}
\item The multiplicity for each sample point is modified by multiplying it by a Gaussian centered around the present state with width $\sigma_k$:
\begin{equation}
m_i^{\prime} = m_i \dfrac{\exp \left[ -\left(\frac{\phi_k - \phi_i}{\sigma_{\phi}}\right)^2 \right]}{N_{\phi}}
\end{equation}

where $N_{\phi}$ is a normalization factor given by:
$$N_{\phi} = \sum_{i=1}^n m_i \exp \left[ -\left(\frac{\phi_k - \phi_i}{\sigma_{\phi}}\right)^2 \right]$$

\item A random sample can now be selected using the following algorithm:

\begin{algorithmic} 
\STATE $U = U(0,1)$
\FOR{$i=1$ to $n$}
  \IF {$m_i^{\prime} < U$}
    \RETURN $\phi_i^{\prime}$
  \ELSE \STATE  $U=U-m_i^{\prime}$
\ENDIF
\ENDFOR
\end{algorithmic}
where U is a random number between 0 and 1.

\item  Since there is no correlation between the $\theta$ and $\phi$ parameters, the posterior probability is calculated as: 
\begin{equation} \label{post}
p = \dfrac{P(D | \theta^{N}, \phi^{N}) \times \Pi(\theta^N) \times \Pi(\phi^N)}{S(\theta^N|\theta^{N-1}) \times S(\phi^N | \phi^{N-1})}
\end{equation}

Here:
\begin{itemize}
\item N is the current state of MCMC and N-1 is the previous state.
\item P is the probability of data given model. In our case, it is the probability of obtaining the calculated spin period given the sampled parameters.
\item $\Pi$ is prior probability.
\item S is transition probability or the stepping function for transitioning from state N-1 to N.
\end{itemize}

\item Using the following simplifications:

\begin{align}
\Pi(\phi^N) & = m_N / \sum_{i=1}^n m_N \\
S(\phi^N | \phi^{N+1}) & = \dfrac{ m_N   e^{  -\left(\tfrac{\phi^{N+1} - \phi^N}{\sqrt{2}\sigma_{\phi}}\right)^2     }}{\sum_i^{n} m_i e^{  -\left(\tfrac{\phi^{N+1} - \phi^{i}}{\sqrt{2}\sigma_{\phi}}\right)^2}} 
\end{align}

, the acceptance ratio reduces to:
\begin{equation} \label{eq:R}
R= \dfrac{P(D | \theta^{N+1},\phi^{N+1})}{P(D | \theta^{N},\phi^{N})}   \dfrac{\Pi(\theta^{N+1})}{ \Pi(\theta^{N})} \dfrac{\sum_{i=1}^{n} m_i \exp \left[  -\left(\tfrac{\phi^{N} - \phi^{i}}{\sqrt{2}\sigma_{\phi}}\right)^2 \right] } {\sum_{i=1}^{n} m_i \exp \left[  -\left(\tfrac{\phi^{N+1} - \phi^i}{\sqrt{2}\sigma_{\phi}}\right)^2 \right]}
\end{equation}

\end{enumerate}

\subsection{Modified MCMC for Fast Convergence}\label{sec:8.2}

Fast convergence of MCMC chains to the target distribution requires fine-tuning of the step size for each parameter.  After running MCMC for several days, we noticed for some of the systems the acceptance rate (i.e., the ratio of the number of accepted steps to the rejected steps after combining all chains) is as low as 1$\%$. For these systems, the computation time for each MCMC iteration is quite large (for some cases, it can go as high as 1 hour). To increase the acceptance rates,  we use the modification to MCMC  following the adaptive tuning idea described in \cite{shaby2010exploring}. Appendix \ref{sec:app} summarized the concept and mathematics of this modification. We apply this to 16 out of 41 binary systems. 

To summarize \cite{shaby2010exploring}, a covariance matrix is calculated after a certain number of steps to find correlations among the parameters. This covariance matrix then serves as a stepping function for the next set of iterations. The covariance matrix keeps updating until an ideal acceptance rate (0.24) is achieved. They proved that this method preserves the Markov property while improving the run time to allow a fast convergence. For our case we apply this method once (instead of periodically updating) for the 16 binary systems with extremely low acceptance rates.

\section*{Data Availability}

The data underlying this article are available in the article and in its online supplementary material.



\bibliographystyle{mnras}
\bibliography{paper} 




\bsp  
\label{lastpage}
\end{document}